# Covalent Nitrogen Doping and Compressive Strain in MoS$_2$ by Remote N$_2$ Plasma Exposure

*Angelica Azcatl[1], Xiaoye Qin[1], Abhijith Prakash[2], Chenxi Zhang[1], Lanxia Cheng[1], Qingxiao Wang[1], Ning Lu[1], Moon J. Kim[1], Jiyoung Kim[1], Kyeongjae Cho[1], Rafik Addou[1], Christopher L. Hinkle[1], Joerg Appenzeller[2] and Robert M. Wallace[1*]*

[1]Department of Materials Science and Engineering, The University of Texas at Dallas, 800 West Campbell Road, Richardson, Texas 75080, United States

[2]Department of Electrical and Computer Engineering, Brick Nanotechnology Center, Purdue University, West Lafayette 47907, Indiana United States

ABSTRACT

Controllable doping of two-dimensional materials is highly desired for ideal device performance in both hetero- and p-n homo-junctions. Herein, we propose an effective strategy for doping of MoS$_2$ with nitrogen through a remote N$_2$ plasma surface treatment. By monitoring the surface chemistry of MoS$_2$ upon N$_2$ plasma exposure using *in-situ* X-ray photoelectron spectroscopy, we identified the presence of covalently bonded nitrogen in MoS$_2$, where substitution of the chalcogen sulfur by nitrogen is determined as the doping mechanism. Furthermore, the electrical characterization demonstrates that p-type doping of MoS$_2$ is achieved by nitrogen doping, in agreement with theoretical predictions. Notably, we found that the presence of nitrogen can induce compressive strain in the MoS$_2$ structure, which represents the first evidence of strain induced by substitutional doping in a transition metal dichalcogenide material. Finally, our first principle calculations support the experimental demonstration of such strain, and a correlation between nitrogen doping concentration and compressive strain in $MoS_2$ is elucidated.

MAIN TEXT

Two-dimensional transition metal dichalcogenides (TMDs) have opened the possibility to scale down the thickness of the semiconductor channel material in field-effect transistors (FETs) to the sub-nanometer regime, while avoiding the dramatic increase in band gap that occurs in 3-D crystals due to the quantum confinement effect.[1] Yet, for the realization of TMD-based devices (e.g. *p-n* diodes, FETs, tunnel FETs), controlled doping type and doping concentration of the TMD material is a key requirement. In this regard, the study of molecular doping using molecules such as $NO_2$,[2] polyethylenimine,[3] and benzyl viologen[4] represents an important step into the understanding of doping TMDs through charge transfer. However due to the non-covalent bonding nature of the dopant molecules, this strategy has been limited by their volatility with time, making the control over doping concentration a challenge. Electrostatic doping is another approach that has been employed to dope TMDs through the use of a polymer electrolyte or ionic liquid gating,[5,6,7] relying on the applied voltage to modulate the carrier density. In addition to these strategies, covalent doping of TMDs, where single atom dopants are introduced in the TMD lattice through metal or chalcogen substitution, is a potential route to achieve stable and controllable doping. An example of covalent p-type doping of $MoS_2$ has been demonstrated through Mo substitution by Nb during the growth process.[8,9] Furthermore, substitutional doping can induce new functionalities to TMD materials such as optical band gap tuning[10] or the prospect of magnetic behavior.[11]

The use of plasma treatments has been proposed as a strategy to incorporate fluorine[12] and phosphorous[13] in $MoS_2$. However, these plasma assisted doping processes can generate undesirable side effects of such as layer etching and degradation of the $I_{ON}$ current for the respective plasma doped $MoS_2$ FETs, which suggests that improvement of the plasma processing for TMDs is still to be developed. Encouraged by the efforts on the exploration of plasma assisted doping strategies for TMDs, the use of remote $N_2$ plasma treatment for the introduction of nitrogen in $MoS_2$ as a dopant atom is investigated in this work. $N_2$ plasma exposure is a practical technique that has been widely used for the incorporation of nitrogen atoms into the lattice of various semiconductors[14,15] and metal gate materials,[16,17] and it has been successfully applied for nitrogen doping of graphene.[18] In the case of monolayer $MoS_2$, nitrogen doping is predicted to induce p-type behavior according to first principles calculations.[19,20] Here, evidence of the covalent nitrogen doping of $MoS_2$ upon remote $N_2$ plasma exposure is provided directly. It was found that a controllable nitrogen concentration can be realized with $N_2$ plasma exposure time. Furthermore, the electrical characterization indicates that nitrogen acts as a p-type dopant in $MoS_2$ and, more importantly, that the electronic performance of the $N_2$ plasma treated $MoS_2$ was preserved in reference with the untreated $MoS_2$. The structural changes associated with the $N_2$ plasma exposure on the $MoS_2$ surface are also presented here. We present the first report of strain induced by a single-atom dopant in a TMD material. Finally, first principles calculations were performed to estimate the relation between strain and nitrogen concentration in $MoS_2$.

The presence and chemical identity of nitrogen in bulk $MoS_2$ upon $N_2$ plasma exposure was investigated by *in-situ* x-ray photoelectron spectroscopy (XPS) and shown in Figure 1a. The as-exfoliated $MoS_2$ (0001) surface was first annealed at 300 °C for one hour under ultrahigh vacuum (UHV) to desorb carbonaceous contamination adsorbed during the short air exposure.[21]

By performing this pre-annealing step, the carbon and oxygen concentrations on the MoS$_2$ surface were below the detection limit of XPS and, consequently, the formation of CN$_x$ species after N$_2$ plasma exposure on the surface was below detectable limits (see Supplementary Figure S2). Then, sequential N$_2$ plasma exposures on the annealed MoS$_2$ surface were carried out. After the first exposure of 2 min ("*t1*"), a low intensity peak at ~ 397.7 eV was detected in the N 1*s* region, indicating the presence of nitrogen in MoS$_2$. It should be noted that the Mo 3$p_{3/2}$ region overlaps the N 1*s* region, as evidenced by a peak associated with MoS$_2$ centered at ~ 394.8 eV. For this reason, the identification of nitrogen related peaks requires a detailed analysis of the high binding energy side of the Mo 3$p_{3/2}$ peak. Examining the Mo 3*d* region, a shoulder at ~0.4 eV higher in binding energy than the Mo-S peak also developed at *t1*. By correlation of the new chemical states in N 1*s* and Mo 3*d*, the formation of a Mo-N bond was identified as represented in Figure 1b, where XPS peak positions are consistent with the literature.[22,23] Additionally, at *t1* the full width of half maximum (FWHM) of both Mo 3*d* and S 2*p* peaks increased by ~25% with respect to as-exfoliated MoS$_2$, reflecting some degree of bond diversity generated at the MoS$_2$ surface. It is noted that nitrogen did not exhibit reactivity with sulfur, as no additional chemical states indicative of S-N bonding (164.8 eV)[24] were detected in the S 2*p* region.

With further plasma exposures, the intensity of the Mo-N bond in the N 1*s* region increased, while no additional nitrogen species were detected throughout the process. In parallel with this increase in nitrogen concentration, the total sulfur content in MoS$_2$ decreased as shown in Figure 1c, suggesting preferential removal of sulfur. These concomitant changes in sulfur concentration with N$_2$ plasma exposure time suggests that the mechanism of formation of N-Mo bonds involves sulfur substitution by nitrogen in the MoS$_2$ structure, consistent with first principle calculations which have shown that nitrogen will behave as a substitutional dopant.[19] The atomic concentration (at. %) of nitrogen in MoS$_2$ achieved by varying the N$_2$ plasma exposure time obtained from the quantitative analysis of the XPS spectra is shown in Figure 1e, where the at% of nitrogen exhibited a logarithmic relation with N$_2$ plasma exposure time. To estimate the areal density of nitrogen atoms on the MoS$_2$ surface, the nitrogen coverage ($\Theta_N$) was calculated from the ratio of the integrated intensities of N 1$s$ to Mo 3$d_{5/2}$, as described in Supplementary section 6 and ref. 25. For the N$_2$ plasma exposure times shown in Figure 1a, the resulting $\Theta_N$ is 0.1 ML for *t1*, 0.35 ML for *t2*, 0.77 ML for *t3*, 0.92 ML for *t4*, and 1.08 ML for *t5*, where 1 ML corresponds to 1.16×10$^{15}$ atoms/cm$^2$ for the MoS$_2$(0001) basal plane. According to these values of nitrogen coverage and angle-resolved XPS measurements (see Supplementary Figure 6S), the introduction of nitrogen by the remote N$_2$ plasma exposure occurs at the outermost MoS$_2$ surface.

Nitridation of MoS$_2$ was previously reported to be achieved by exposing MoS$_2$ powder to NH$_3$ at 750 °C. Another approach for nitrogen doping of MoS$_2$ was reported to be performed by synthesis of MoS$_2$ nanosheets using a sol-gel method.[26] In contrast, the advantage of the N$_2$ plasma process described in this work is that a controllable concentration of the nitrogen dopant was realized, where the newly formed Mo-N bond owns a covalent character due to the strong hybridization between the N 2$p$ and Mo 4$d$ orbitals.[27,28] In fact, good thermal stability of the Mo-N bond and lack of N desorption was confirmed upon annealing at 500 °C (see Supplementary Figure S4).

According to the peak positions for Mo 3$d$ and S 2$p$, the initial as-exfoliated MoS$_2$ surface exhibited the commonly observed unintentional n-type doping.[29] Interestingly, all regions in the XPS spectra shifted identically to lower binding energies with N$_2$ plasma exposure from *t1* to *t3* as shown in Figure 1c, suggesting a change in the Fermi level associated with p-type doping.[29] Yet, the interpretation of the magnitude of the Fermi level shift for nitrogen doped MoS$_2$ (N-doped MoS$_2$) is not straightforward. The charge transfer due to the presence of nitrogen as a p-type dopant,[20] with the band bending induced by the formation of Mo-N covalent bonds at the top-most layer and the preferential sulfur removal,[30] will both contribute to the measured shift. The overall effect on the valence band spectra reveals that the Fermi level moves closer to the MoS$_2$ valence band.

A comparative band alignment of MoS$_2$ before and after 15 min of N$_2$ plasma treatment is shown in Figure 2a. The band diagrams were constructed from XPS measurements (see Supplementary Figure S5), and employ a band gap value of 1.23 eV ± 0.02 eV for bulk MoS$_2$,[31] which is assumed to remain constant upon nitrogen doping. Figure 2a shows that the work function (Φ) increases after the N$_2$ plasma treatment, and the valence band maximum (VBM) shifted from 0.87 eV below the Fermi level for as-exfoliated MoS$_2$ to 0.34 eV for nitrogen doped MoS$_2$, indicating a Fermi level shift that provides evidence of p-type doping. It is also noted that during the doping process described in this work, the oxygen and carbon concentration on the MoS$_2$ surface were below XPS detection limits (see Supplementary Figure 3S). Therefore, nitrogen was exclusively the only element involved in the Fermi level shift towards the valence band in MoS$_2$. Finally, the estimated electron affinity decreased after nitrogen doping, which can be related to the modification of surface termination in MoS$_2$, from sulfur to nitrogen terminated.[32]

To further investigate the electrical properties of the nitrogen doped MoS$_2$ and support the initial evidence of p-type doping given by XPS, back gated field effect transistors (FETs) were fabricated with the device structure as shown in Figure 2c. For the electrical characterization, a 15 min $N_2$ plasma exposure was performed on $MoS_2$ flakes with different thicknesses exfoliated on a $SiO_2$/Si substrate. Figure 2b shows the $I_{DS}$-$V_{GS}$ characteristics for a representative back gated nitrogen doped multilayer $MoS_2$ FET. The positive shift of the threshold voltage $V_{th}$ is consistent with the expected p-type dopant behavior of nitrogen in $MoS_2$.[33] Yet, a p-type branch in the characteristics was not observed, probably due to well-known Fermi level pinning of the metal contacts the near the $MoS_2$ conduction band.[34] Interestingly, and in contrast to Nb[35] or $NO_2$[2] doping, no degenerate doping was observed after $N_2$ plasma exposure, which highlights the advantage of this process as the control of the device current with the applied voltage was preserved. Furthermore, when correcting for the aforementioned threshold voltage shift, device characteristics in the device ON-state above threshold voltage reached almost identical on-current levels $I_{ON}$, which suggests that the presence of nitrogen did not enhance scattering so as to reduce the $I_{ON}$ of the transistors.

A dependence of the $V_{th}$ shift with layer thickness was also found, as shown in Figure 2d, where $V_{th}$ increased with decreasing $MoS_2$ layer thickness. The higher $V_{th}$ shift for mono and few-layer $MoS_2$ is consistent with an expected higher relative atomic concentration for nitrogen, as compared with thick $MoS_2$. For thicker $MoS_2$ (>6 layers), the underlying $MoS_2$, which is expected to be nitrogen free based on the XPS analysis, will produce the more dominant electrical response. From these electrical characteristics, the p-type doping level ($N_A$) calculated for the nitrogen doped $MoS_2$ ranges from $2.5 \times 10^{18}$ cm$^{-3}$ to $1.5 \times 10^{19}$ cm$^{-3}$, based on a reference value of $N_A \sim 1.5 \times 10^{18}$ cm$^{-3}$ for the as-exfoliated $MoS_2$.[33] To the best of our knowledge, this is the first report of an estimation of a substitutional acceptor dopant concentration for nitrogen doped $MoS_2$. Here, it is worth noting that the XPS measurements indicate that the initial bulk $MoS_2$ was unintentionally n-type doped, of which the outermost layer was then counter-doped upon $N_2$ plasma exposure, as shown in figure 2a. In the case of the exfoliated $MoS_2$ used for the electrical characterization presented here, such flakes exhibit an initial p-type doping possibly due to water intercalation,[33] and after nitrogen doping, an increase in the acceptor doping level was generated. Therefore, both measurements support the claim that nitrogen acts as a p-type dopant for $MoS_2$, which is consistent with the theoretical predictions.[19,20]

The chemical characterization indicates that the $N_2$ plasma treatment is an effective method for nitrogen doping of $MoS_2$. However, the use of plasma exposures on the $MoS_2$ can result in side-effects such as layer etching[12,13,36] or roughening of the surface.[37,38] To evaluate the effect of the $N_2$ plasma treatment on the $MoS_2$ surface topography, atomic force microscopy (AFM) imaging was employed. The initial surface of a $MoS_2$ flake exfoliated on $SiO_2$/Si is shown in Figure 3a, where a step height of around 6 nm was observed for this flake. After 15 min of $N_2$ plasma exposure, no evidence of layer etching was detected as the height of the step in the flake remained constant, and no additional features which indicate physical damage were present, such as cluster-like particles or etched areas. Instead, the $MoS_2$ surface remained smooth with a root mean square (RMS) roughness of 0.74 nm.

Interestingly, increasing the $N_2$ plasma exposure time to 60 min causes the development of cracks across the sample surface primarily at the grain boundaries of the $MoS_2$ surface. The depth of the cracks was consistently ~0.7 nm, suggesting that the crack formation occurred just within the top most, N-doped $MoS_2$ monolayer. The majority of the cracks were connected at an angle of ~120°, where the crack propagation reflected a hexagonal-like symmetry. Interestingly, there was a minimal effect on the surface roughness, indicating that a longer plasma exposure did not cause disruption of the underlying $MoS_2$ layered structure. Further evidence of the preservation of the layered structure in $N_2$ plasma treated $MoS_2$ is provided by cross-section STEM images, shown in Figure 3b. The contrast change observed, due to the presence of nitrogen in $MoS_2$, indicates that the sulfur substitution occurred predominantly within the outermost $MoS_2$ layer, in agreement with the XPS analysis. Since nitrogen has a smaller atomic radius in comparison to sulfur, its introduction to the $MoS_2$ lattice and subsequent formation of Mo-N bonds could potentially generate compressive strain, resulting in development of cracks for a high nitrogen concentration at the surface. To investigate this hypothesis, Raman Spectroscopy on few-layer, exfoliated $MoS_2$ flakes were carried out.

Figure 3c shows that the Raman spectra for an exfoliated $MoS_2$ flake on $SiO_2$/Si substrate. The peak separation between the $A_{1g}$ and $E^1_{2g}$ vibrational modes is 21.5 cm$^{-1}$, which according to the literature corresponds to exfoliated, bilayer $MoS_2$.[39] After 15 min of the $N_2$ plasma treatment, the $E^1_{2g}$ peak corresponding to an in-plane vibrational mode blue shifted and split into two peaks, labeled as $E^{1+}_{2g}$ as $E^{1-}_{2g}$. The $E^1_{2g}$ peak splitting has been correlated to symmetry breaking of this vibrational mode generated due to the presence of strain in the $MoS_2$ layered structure.[40] The blue shift of the $E^1_{2g}$ mode with respect to the initial exfoliated $MoS_2$ suggests that that the type of strain generated due to the presence of Mo-N bonding is compressive,[41] which suggests contraction of the $MoS_2$ lattice due to the introduction of nitrogen as a dopant. To the best of our knowledge, this is also the first report of strain induced by a single atom dopant in two dimensional TMDs. In fact, this situation is analogous to the case of silicon, where compressive strain is generated due to doping with boron or carbon atoms.[42,43] In addition to the effect of compressive strain for the N-doped $MoS_2$ system, charge doping effects are likely to contribute to the measured Raman signal, which is translated as a shift of the out-of-plane vibrational mode $A_{1g}$.[44] Consistent with the p-type doping of $MoS_2$ given by substitutional nitrogen, the $A_{1g}$ peak blue shifted by 0.43 ± 0.07 cm$^{-1}$ after 15 min of $N_2$ plasma exposure.

Increasing the $N_2$ plasma exposure time to 60 min caused a blue shift up to 6.4 cm$^{-1}$ for $E_{2g}^{1+}$ and 1.2 cm$^{-1}$ for $A_{1g}$ with respect to the initial peak positions, as shown in Figure 3c. Additionally, the $E_{2g}^1$ peak intensity was significantly reduced, where the $E_{2g}^1$ to $A_{1g}$ ratio of the respective integrated intensities decreased by 47% with respect to as-exfoliated $MoS_2$. The decrease in the $E_{2g}^1/A_{1g}$ ratio is an indication of suppression of the in-plane movement related to the $E_{2g}^1$ mode, which results from compressive strain in the $MoS_2$ structure according to studies on $MoS_2$ subjected to externally applied pressure.[45] Interestingly, there was a decrease in the FWHM of the $A_{1g}$ feature from 4.5 cm$^{-1}$ for exfoliated $MoS_2$ to 3.7 cm$^{-1}$ after the 60 minute exposure. According to the dependence of $A_{1g}$ FWHM on the $MoS_2$ thickness,[46] this decrease in FWHM is an indication of thinning down the effective $MoS_2$ thickness, from bilayer to monolayer due to a complete sulfur substitution by nitrogen in the top most $MoS_2$ layer. At this stage, the effect of strain is still present in the remaining bottom $MoS_2$ layer, which is evidenced by the fact that the $A_{1g}$ and $E_{2g}^1$ peaks are blue shifted with respect to the peak positions reported for unstrained monolayer $MoS_2$.[39] Estimation of compressive strain from the $E_{2g}^1$ shift based on models developed for biaxial strain induced by external forces[47] is not feasible for this N-doped bilayer $MoS_2$ system since the shifts in the Raman spectra are the convolution of compressive strain and charge doping effects due to sulfur substitution by nitrogen.

In an attempt to estimate the magnitude of strain at different nitrogen concentrations in $MoS_2$, first principles calculations were performed for the N-doped bilayer $MoS_2$ system. For these calculations, a 4×4 supercell (12.65 Å × 12.65 Å) of bilayer $MoS_2$ was used. When a sulfur atom is substituted by a nitrogen atom in $MoS_2$, the resulting bond length for the covalent Mo-N bond is 2.01 Å whereas that of Mo-S in pristine $MoS_2$ is 2.41 Å. Therefore, the presence of Mo-N generates a contraction of the $MoS_2$ lattice, which is in agreement with Raman Spectroscopy measurements. Figure 4 shows the percentage of strain for different nitrogen coverages. The magnitude of strain was calculated as $\varepsilon = (a_N - a_p)/a_p$ where $a_N$ and $a_p$ (=3.16 Å) are the lattice constant of N-doped $MoS_2$ and pristine $MoS_2$, respectively. By correlation of the calculated dependence of compressive strain with N coverage with the experimental XPS data shown earlier, the estimated compressive strain for the different N coverages would correspond to: *t1*= 0.1%, *t2*= 0.5%, *t3*= 1.3%, *t4*= 1.5% and *t5*= 1.7%. According to these calculations, the magnitude of the strain when nitrogen is present as dopant in $MoS_2$ is comparable to that obtained from mechanically induced strain,[48] suggesting that this approach can potentially be applied to tune the optical band gap of $MoS_2$[40]. Based on the DFT calculations, the percentage of compressive strain given by one monolayer coverage of nitrogen in $MoS_2$ is 1.7%. Experimentally, this amount of strain would correspond to a 60 min $N_2$ plasma exposure, after which cracking of $MoS_2$ occurred. It has been reported that the strain required to cause breaking of the $MoS_2$ structure by mechanical forces is in the range of ~6-11%.[49] Therefore, the estimation of strain given by the presence of Mo-N bonds suggests that the cracking process observed in the outermost $MoS_2$ layers is not only related to the compressive strain, but also to other factors that can promote crack formation. Recent theoretical calculations have shown that the amount of strain needed to cause cracking of $MoS_2$ can be reduced by chemical absorption of molecules on the $MoS_2$ surface, since the generation of crack lines results energetically favorable at the adsorption sites during chemisorption.[50] Therefore, the chemical adsorption of nitrogen possibly promotes the cracking process by reducing the amount of strain needed to generate crack lines. Importantly, the experimental evidence of strain induced by single atom doping presented here opens the possibility to identify the window at which doping concentration and strain can be used to modify the electronic properties of two-dimensional TMDs.

Finally, it should be noted that given the evidence of compressive strain generated in $MoS_2$ due to nitrogen doping, a band gap change of $MoS_2$ is not excluded. If the $MoS_2$ band gap was modified due to strain, that change is expected to have a minimal contribution in the electrical characteristics presented here. The percentage of strain for the nitrogen doped $MoS_2$-based FET measurements is estimated to be of the order of ~1.3%. According to previous reports, such levels of strain, generated mechanically, can induce a band gap change on the order of ~100 meV or less depending on the $MoS_2$ layer thickness.[48] While in general a change in bandgap can be extracted from changes in the device characteristics,[51] the absence of clear ambipolar behavior in the case of $MoS_2$ prevents such an analysis here. However, using the inverse subthreshold slope from Figure 2b to translate the gate voltage axis into an energy scale, we conclude that a hypothetical increase in the $MoS_2$ band gap of 100 meV would only explain a change in the threshold voltage of up to 2 V, much smaller than the observed threshold voltage shifts. Therefore the $V_{th}$ shift shown in this work is expected to be dominated by charge doping given by nitrogen. Further studies will be needed to determine possible strain induced band gap tuning of $MoS_2$ upon nitrogen doping, in a controllable manner.[48]

In summary, the use of $N_2$ plasma as a strategy for nitrogen doping of $MoS_2$ was evaluated in this work. The surface chemistry of $MoS_2$ upon $N_2$ plasma treatment indicated that nitrogen forms a covalent bond with molybdenum through chalcogen substitution of S. It was found that the nitrogen concentration in $MoS_2$ can be controlled with $N_2$ plasma exposure time. The electrical characterization shows that nitrogen acts as substitutional p-type dopant in $MoS_2$, consistent with theoretical predictions and XPS analysis, while the electrical performance of the nitrogen doped $MoS_2$ based FETs was preserved in reference to the as-exfoliated $MoS_2$ based FETs. It was also demonstrated that the doping process through the use of remote $N_2$ plasma induced compressive strain in $MoS_2$, and while the layered structure of $MoS_2$ was not damaged, cracking did occur with significant concentrations of N incorporated. It was shown that nitrogen concentration can be applied to tune the level of compressive strain in $MoS_2$. This work paves the way for the realization of substitutional covalent doping in two dimensional materials.

FIGURE CAPTIONS

**Figure 1.** Nitrogen doping of $MoS_2$ through sulfur substitution. (a) XPS spectra from as-exfoliated $MoS_2$ showing the N 1$s$, Mo 3$d$ and S 2$p$ core levels after annealing and after sequential $N_2$ plasma exposures, where $t1$= 2 min, $t2$=7 min, $t3$=15 min, $t4$=30 min, $t5$=60 min. (b) Schematic of the covalent nitrogen doping in $MoS_2$ upon $N_2$ plasma surface treatment. (c) Stoichiometry for the N-doped $MoS_2$ system represented as $N_xMoS_y$. (d) peak positions for Mo 3$d$ and S 2$p$ from $MoS_2$ with respect to $N_2$ plasma exposure time obtained from the XPS spectra in (a). (e) Atomic percentage (at %) of nitrogen in $MoS_2$ as a function of $N_2$ plasma exposure time.

**Figure 2.** p-type doping effect of nitrogen in $MoS_2$. (a) Energy band diagram for as-exfoliated $MoS_2$ and nitrogen doped $MoS_2$, where the work function ($\Phi$) and the valence band maximum (VBM) were acquired from XPS measurements from the secondary electron cut-off energy and the valence band edge, respectively. The band gap ($E_g$) was assumed to be that of bulk $MoS_2$ (1.23 eV)[31] for both cases. The measured $\Phi$ and VBM values were employed to estimate the electron affinity ($\chi$). All values shown here are in eV units. (b) $I_{DS}$-$V_{GS}$ characteristics of multilayer nitrogen doped MoS$_2$ FET. (c) Schematic of the back-gated nitrogen doped MoS$_2$ FET on Si/SiO$_2$ structure used in this study. (d) Dependence of $V_{th}$ on the MoS$_2$ layer thickness, lines drawn to guide the eye.

**Figure 3.** Surface topography and van der Waals structure of N-doped MoS$_2$, and compressive strain in MoS$_2$ induced by nitrogen doping (a) Atomic force microscopy images of the surface topography of a MoS$_2$ flake deposited on a SiO$_2$/Si substrate, as-exfoliated and after 15 min and 60 min of N$_2$ plasma exposure. The line profile in blue extends across a MoS$_2$ step, which height is shown in the bottom graphs. (b) Cross-section STEM images of a N$_2$ plasma treated MoS$_2$ samples with exposure time of 60 min, showing the preservation of the layered structure. (c) Raman spectra from a bilayer MoS$_2$ flake deposited on a SiO$_2$/Si substrate, as-exfoliated and after a sequential N$_2$ plasma exposure of 15 min and 60 min. Raman shift *vs* N$_2$ exposure time obtained from the measurements shown in a.

**Figure 4.** Theoretical estimation of the magnitude of compressive strain for N-doped MoS$_2$. Compressive strain dependence on N coverage on bilayer MoS$_2$ obtained from DFT calculations, where N is present a substitutional dopant, where the top view of the optimized N-doped MoS$_2$ structure shows the resulting Mo-N bond length.

# Figure 1

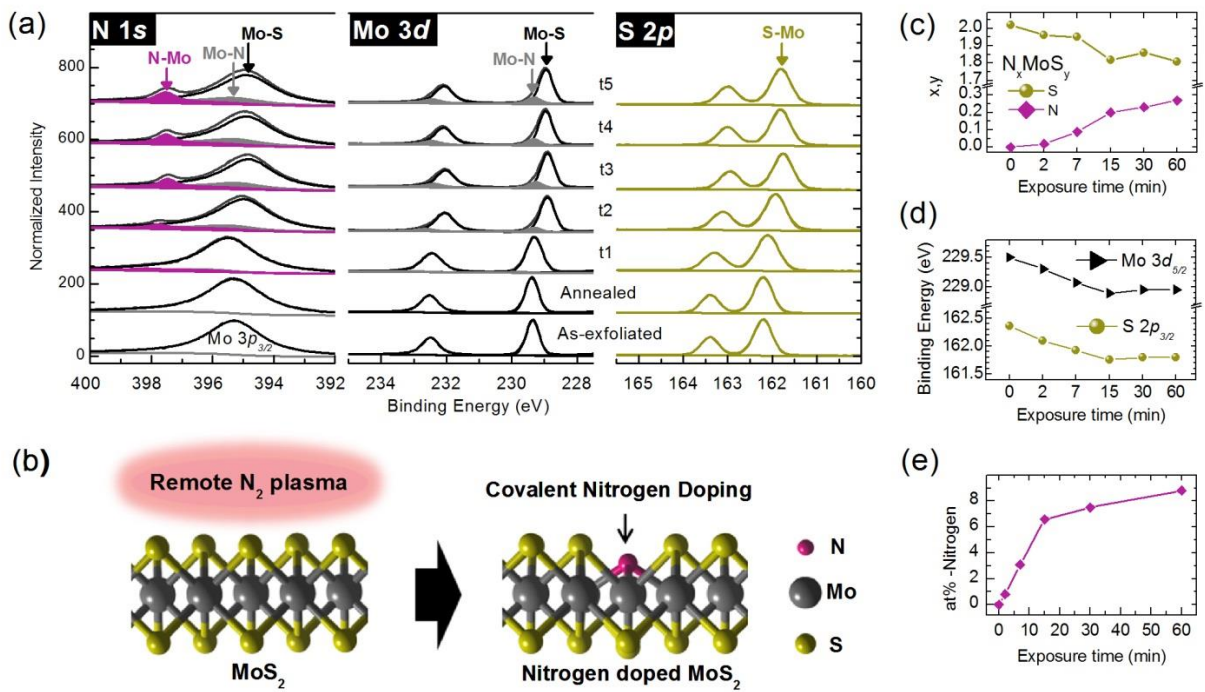

# Figure 2

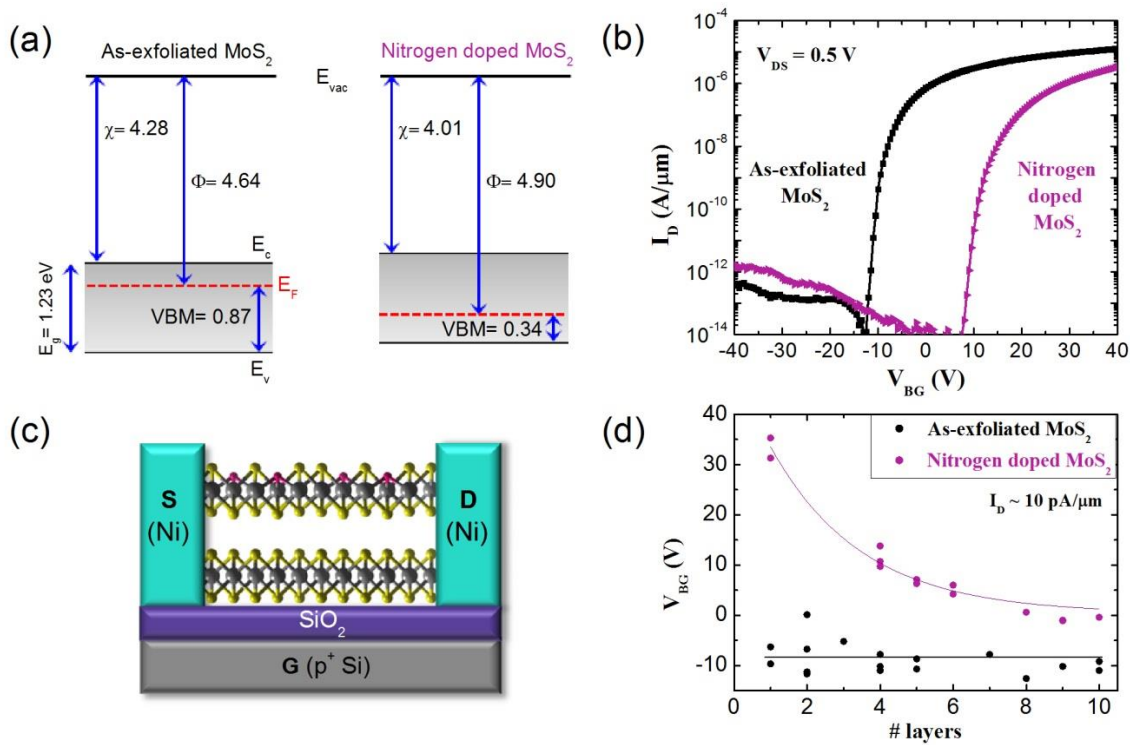

**Figure 3**

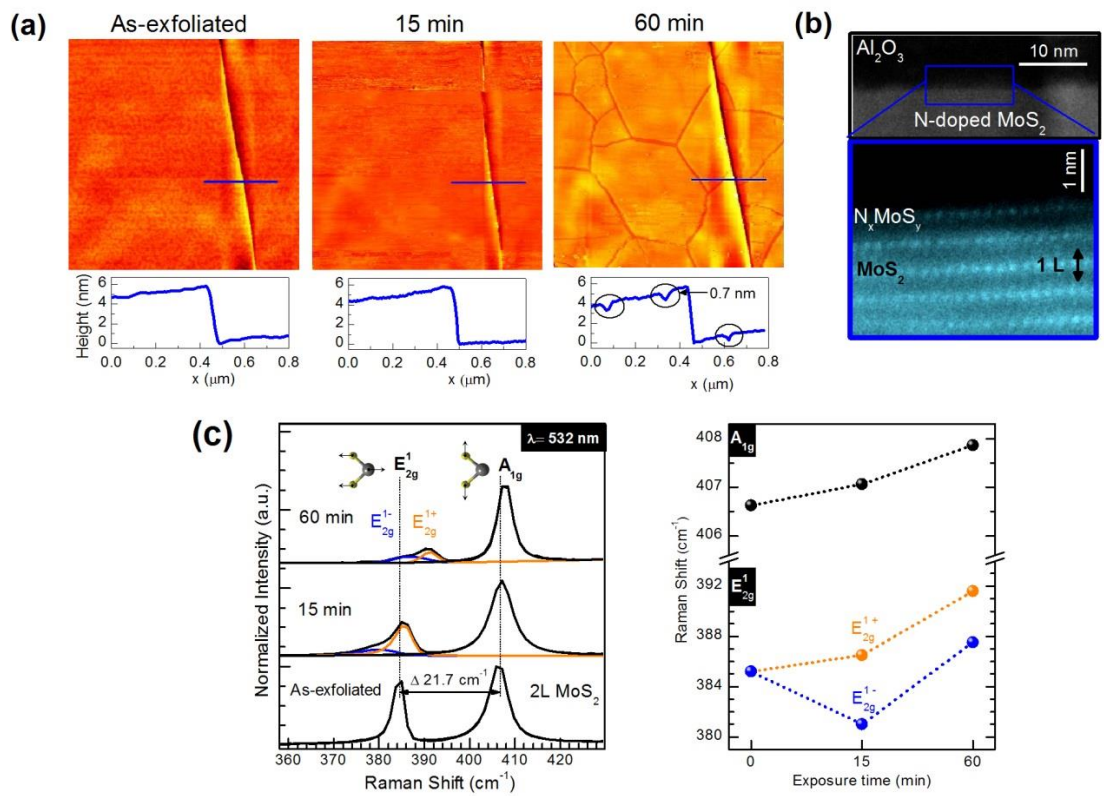

**Figure 4**

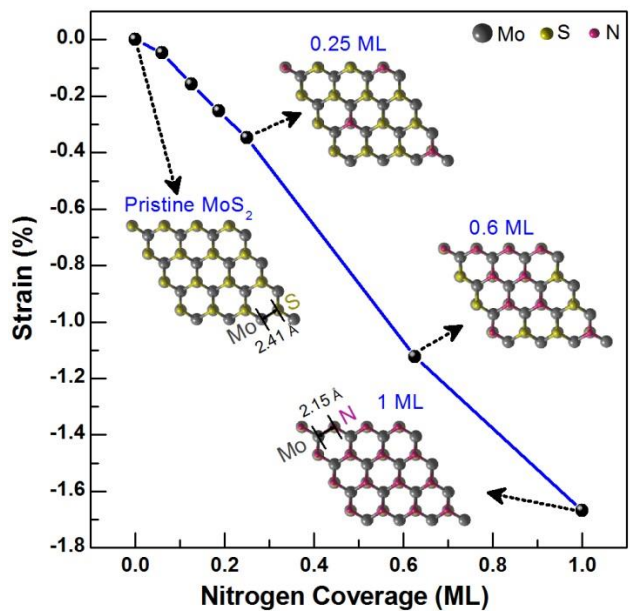

## ASSOCIATED CONTENT

**Supporting Information**.

The experimental methods, O 1$s$ and C 1$s$ XPS spectra for MoS$_2$ upon sequential N$_2$ plasma exposure, the thermal stability of the N-Mo bond, the XPS spectra employed to construct the band diagrams, and description of nitrogen coverage calculation. This material is available free of charge via the Internet at http://pubs.acs.org.

## AUTHOR INFORMATION

**Corresponding Author**

*Email: rmwallace@utdallas.edu

**Notes**

The authors declare no competing financial interest.

## ACKNOWLEDGMENT

The authors thank Dr. William Vandenberghe for useful discussions. This work is supported in part by the Center for Low Energy Systems Technology (LEAST), one of six centers supported by the STARnet phase of the Focus Center Research Program (FCRP), a Semiconductor Research Corporation program sponsored by MARCO and DARPA. It is also supported by the SWAN Center, a SRC center sponsored by the Nanoelectronics Research Initiative and NIST.

# *Supplementary Information*

*Angelica Azcatl[1], Xiaoye Qin[1], Abhijith Prakash[2], Chenxi Zhang[1], Lanxia Cheng[1], Qingxiao Wang[1], Ning Lu[1], Moon J. Kim[1], Jiyoung Kim[1], Kyeongjae Cho[1], Rafik Addou[1], Christopher L. Hinkle[1], Joerg Appenzeller[2] and Robert M. Wallace[1*]*

[1]Department of Materials Science and Engineering, The University of Texas at Dallas, 800 West Campbell Road, Richardson, Texas 75080, United States

[2]Department of Electrical and Computer Engineering, Brick Nanotechnology Center, Purdue University, West Lafayette 47907, Indiana United States

**Outline**

1. Experimental methods

2. $N_2$ plasma exposure on unannealed $MoS_2$

3. O 1*s* and C 1*s* spectra of as-exfoliated $MoS_2$ upon sequential $N_2$ plasma exposure

4. Thermal stability of N-Mo bond

5. Band alignment changes upon $N_2$ plasma exposure

6. Nitrogen coverage on $N_2$ plasma treated $MoS_2$ by XPS

# 1. Experimental methods

Nitrogen doping of MoS$_2$ was achieved by exposing the MoS$_2$ surface to remotely generated N$_2$ plasma using a 13.56 MHz RF plasma at a power of 100 W, using N$_2$ as gas source flowing at 45 sccm, at a substrate temperature of 300 °C. All the remote plasma exposures were performed in a chamber which has a base pressure of ~$10^{-9}$ mbar and an operating pressure kept at $7\times10^{-3}$ mbar during the remote plasma exposure. The distance from the source to the sample surface is ~30 cm. *In-situ* X-ray photoelectron spectroscopy characterization was performed in an ultrahigh vacuum (UHV) system described elsewhere.[1] XPS was carried out using a monochromated Al Kα source (hν = 1486.7 eV) and an Omicron EA125 hemispherical 7-channel analyzer. The XPS scans were acquired at a take-off angle of 45° with respect to the sample normal and pass energy of 15 eV. For XPS peak analysis and deconvolution, the software *AAnalyzer* was employed, where Voigt line shapes and an active Shirley background were used for peak fitting.[2] For the *in-situ* XPS characterization, natural bulk MoS$_2$ (SPI) was mechanically exfoliated using Scotch™ tape to peel-off the top-most surface layers, followed by loading into the UHV system (within five minutes). As noted in the text, annealing in UHV was conducted at 300 °C for one hour to desorb physisorbed species from the minimal air exposure of the exfoliated surface.

For the electrical characterization, MoS$_2$ flakes of various thicknesses were mechanically exfoliated onto two identical substrates, each with a top layer of 90 nm SiO$_2$ and underlying heavily doped (p++) Si. Flakes of various thicknesses were initially approximately identified by optical contrast and the flake thicknesses were later determined accurately by atomic force microscopy (AFM). MoS$_2$ flakes on one of the two substrates were subjected to annealing, followed by the N$_2$ plasma treatment mentioned in the main text. Both sets of samples, one set consisting of as-exfoliated MoS$_2$ and the other consisting of N$_2$ plasma treated MoS$_2$ flakes were processed together to fabricate FETs. Electron beam lithography, followed by electron beam evaporation and lift-off was used to define the source and drain contacts. Ni was used as the contact metal. After fabrication of the devices, the electrical characteristics were obtained in air in a Lakeshore probe station using an Agilent semiconductor parameter analyzer. In all measurements, the 90 nm SiO$_2$ was used as the gate dielectric and the highly doped Si underneath was used as the back gate.

For Raman Spectroscopy, $MoS_2$ flakes were transferred onto 300 nm $SiO_2$/Si substrate by mechanical exfoliation. The Raman spectra were acquired using a 532 nm wavelength laser using a Renishaw confocal Raman system model inVia under air-ambient conditions. AFM characterization was performed on tapping mode, using an Atomic Probe Microscope (Veeco, Model 3100 Dimension V), also under air-ambient conditions. The processing of the AFM images was performed using WSxM 4.0 software. High-resolution transmission electron microscopy (TEM) cross-sectional specimens of the N-doped $MoS_2$ samples were made by a FEI Nova 200 dual-beam focussed ion beam (FIB)/scanning electron microscope (SEM) with the lift-out method. In FIB, $SiO_2$ and Pt layers were deposited to protect the interested region during focused Ga ion beam milling. For TEM imaging, a JEOL ARM200F operated at 200 kV with a probe aberration corrector was used for atomic resolution high angle angular dark field (HAADF) scanning transmission electron microscopy (STEM). The contrast in a HAADF-STEM image is approximately proportional to $Z^2$, where Z is the atomic number.[3] The modeling of the nitrogen-doped bilayer $MoS_2$ is performed within the framework of density functional theory (DFT) using Vienna *ab initio* package (VASP).[4] Calculations based on the generalized gradient approximation (GGA) using the Perdew-Burke-Ernzerhof (PBE)[5] functional are carried out with projector augmented wave (PAW)[6] pseudopotential plane-wave method. The few-layer, N-doped $MoS_2$ structures are optimized with a vacuum thickness of about 18 Å. The Monkhorst-Pack k-point sampling method in Brillouin zone is Γ-centered with a 4×4×1 mesh in ionic optimization.[7] The cutoff energy is 500 eV and the criteria of convergence for energy and force are set to be $1\times10^{-4}$ eV and 0.02 eV/Å, respectively.

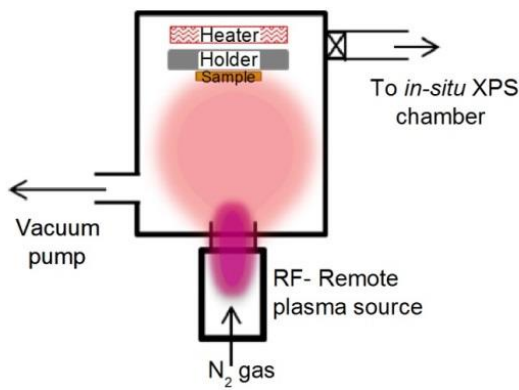

**Figure 1S.** Schematic of the remote plasma system employed for the $N_2$ plasma exposures on $MoS_2$.

## 2. N$_2$ plasma exposure on unannealed MoS$_2$

In this study it was found that a pre-annealing step to further clean the as-exfoliated MoS$_2$ surface of physisorbed species is critical for the formation of covalent bonding between nitrogen and molybdenum in MoS$_2$. As shown in Figure 1S, two peaks in the N1$s$ region are detected at 398.6 eV and 400.1 eV on the unannealed MoS$_2$ surface after N$_2$ plasma exposure. By correlation with the C 1$s$ region, these chemical species are identified as CN$_x$ and CN$_x$-H, respectively.[8] Therefore, without the pre-annealing step, the adsorbed carbon that is present on the initial MoS$_2$ surface (likely physisorbed from the brief atmospheric exposure after exfoliation and prior to placement in vacuum) is readily available to react with the nitrogen species present in the remote N$_2$ plasma to form the aforementioned cyano species, complicating the quantitative analysis of the N 1$s$ region for the identification of the Mo-N bond.

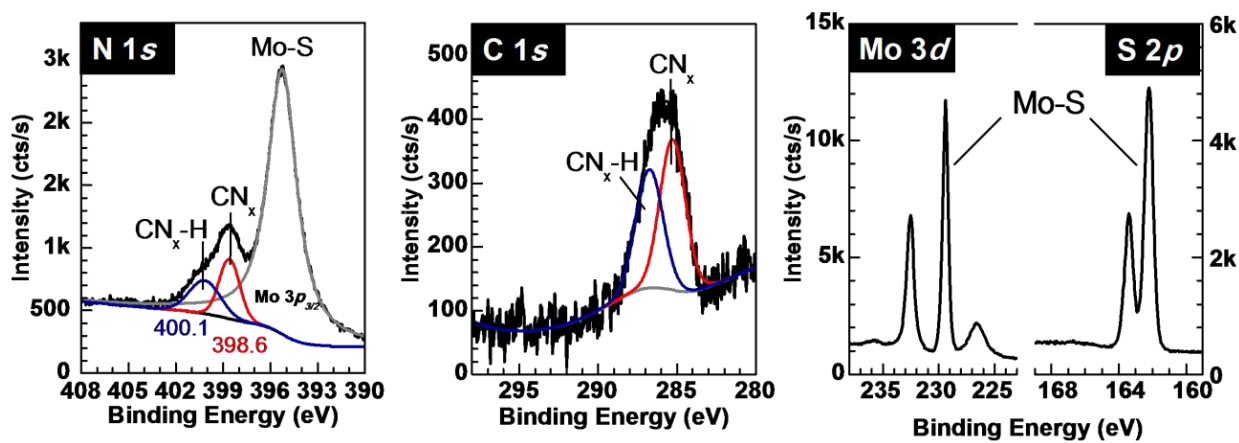

**Figure 2S.** XPS spectra from N 1$s$, C 1$s$, Mo 3$d$ and S 2$p$ regions from as-exfoliated MoS$_2$ with no pre-annealing step after N$_2$ plasma treatment exposed for 30 minutes.

## 3. O 1$s$ and C 1$s$ spectra of as-exfoliated MoS$_2$ upon sequential N$_2$ plasma exposure

The C 1$s$ and O 1$s$ spectra from the sequential N$_2$ plasma exposure on MoS$_2$ described in the main text are shown in Fig. 2S. On the initial as-exfoliated MoS$_2$, a peak corresponding to adsorbed carbon is identified in the C 1$s$ region, while no oxygen signal was detected. By preforming an annealing step at 300 C under UHV for one hour, the signal from the adsorbed carbon was reduced to below detection limit. After sequential N$_2$ plasma exposures on the MoS$_2$ surface, the carbon and oxygen signals remained below detection limit throughout the process. This situation highlights the fact that there was no contribution from oxygen or carbon for the quantitative analysis presented in this work. Therefore, undesirable by-products (e.g. oxy molybdenum nitride, or $CN_x$) were not formed on the $MoS_2$ surface, which was achieved by performing the $N_2$ plasma exposures and XPS analysis *in-situ* in a UHV environment.

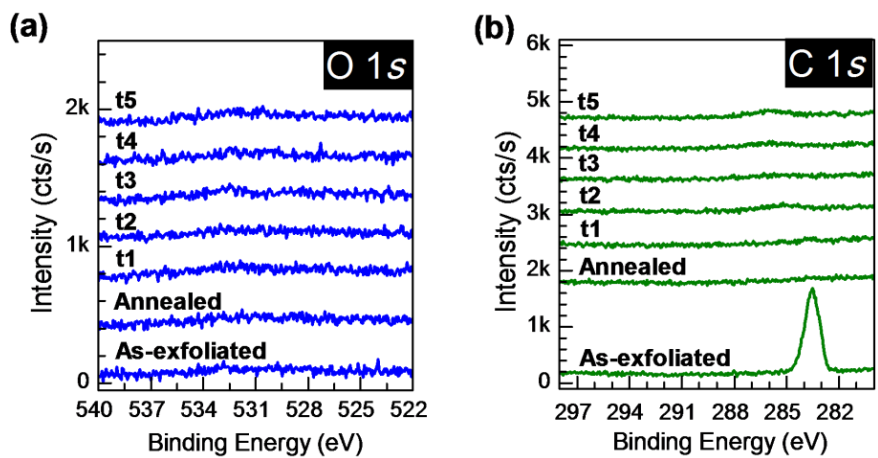

**Figure 3S.** (a) O 1*s* and (b)C 1*s* from as-exfoliated $MoS_2$, after annealing at 300 C for one hour and after sequential $N_2$ plasma exposures (*t1*=2 min, *t2*=7 min, *t3*=15 min, *t4*=30 min, *t5*=60 min).

## 4. Thermal stability of N-Mo bond

To test the thermal stability of the Mo-N bond present on the $MoS_2$ surface, annealing of the $N_2$ plasma-treated $MoS_2$ was performed at 300 °C and 500 °C under UHV conditions (~$10^{-9}$ mbar). As seen in Fig. 3S, it was found that the N-Mo/Mo-$S_{(Mo3p_{3/2})}$ ratio of 0.072±0.004 remained constant upon annealing, which suggests that no desorption or thermal diffusion of nitrogen occurred at these temperatures. The thermal stability of the N-Mo bond upon annealing is consistent with its covalent bonding nature.

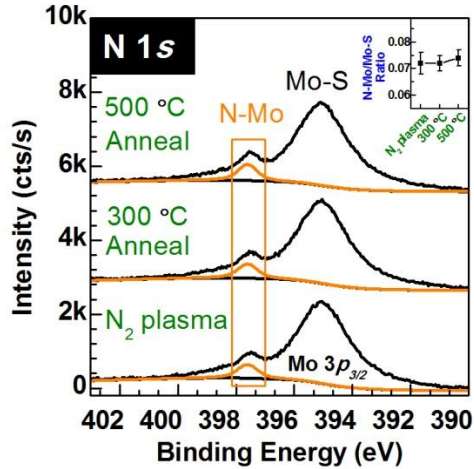

**Figure 4S.** N 1$s$ XPS spectra showing the N-Mo bond in MoS$_2$ generated after N$_2$ plasma and after annealing for hour at 300 °C and 500 °C under UHV.

## 5. Band alignment changes upon N$_2$ plasma exposure

As described in this work, the core levels shifted to lower binding energies with respect to the initial peak positions for as-exfoliated MoS$_2$. To further investigate the changes in the band alignment caused by the sulfur substitution by nitrogen upon N$_2$ plasma exposure, the work function Φ and the valence band maximum (VBM) were obtained from the secondary electron cutoff energy and the valence band edge respectively, both measured by XPS. Figure 4S shows that after 15 min of N$_2$ plasma exposure, the work function increased by ~0.26 eV suggesting that N doping can potentially be applied to increase the work function of MoS$_2$. In addition, the Fermi level shift caused by nitrogen doping moves the valence band maximum from 0.87± 0.13 eV corresponding to n-type MoS$_2$ to a value of 0.34± 0.07 eV for p-type doping.

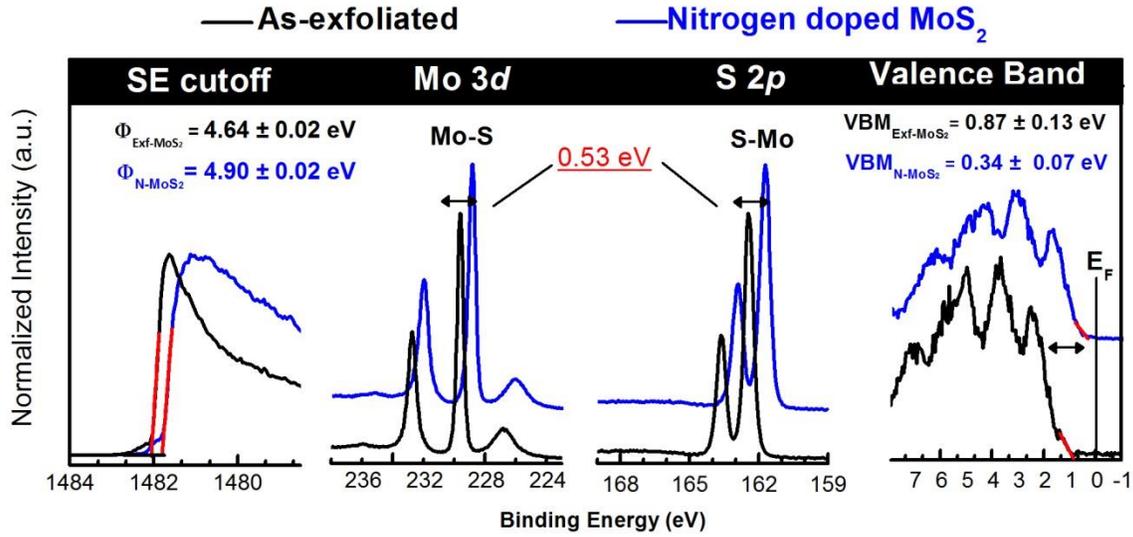

**Figure 5S.** XPS measured work function (Φ) and valence band maximum (VBM) of as-exfoliated and N-doped $MoS_2$ obtained from the secondary electron cutoff energy and the valence band edge, respectively. Error bars for the linear fitting of the secondary electron cut-off energy and valence band edge are shown.

## 6. Nitrogen coverage on $N_2$ plasma treated $MoS_2$ by XPS

In this work, XPS spectra was used to estimate the nitrogen coverage ($\Theta_N$) on $MoS_2$ according to the procedure reported by X. Qin, et al.[9] and V. M. Bermudez, et al.[10], using the following equation:

$$\Theta_N = \frac{I_N}{S_N} \bigg/ \frac{I_{Mo}}{S_{Mo} \sum_{n=0}^{\infty} \exp[-n\, d_{MoS_2}/\lambda_{Mo} \sin(\theta)]}$$

Where $I_N$ and $I_{Mo}$ are the integrated intensities from the N-Mo peak in N 1s and the Mo-S peak in Mo $3d_{5/2}$, respectively, $\lambda_{Mo}$ is the inelastic mean free path for Mo 3d core level electrons in $MoS_2$ obtained from the NIST electron EAL Database, $n$ corresponds to the number of planes that contributes to the XPS signal, $d_{MoS_2}$ = 0.62 nm is the distance between two molybdenum planes in the MoS$_2$ structure, and $S_{Mo}$ and $S_N$ are the relative sensitivity factors for N 1$s$ and Mo 3$d_{5/2}$ which values correspond to 1.731 and 0.477, respectively.[11] All XPS spectra shown in the main text were acquired at a take-off angle of 45° (angle between the sample surface plane and the detector).

To further study the depth at which nitrogen is present in MoS$_2$, a bulk-sensitive take-off angle of 80° was also used to acquire the XPS data after 60 min N$_2$ plasma treatment on MoS$_2$. Figure 5S shows that the N-Mo peak intensity decreased when the spectra was acquired at the 80° take-off angle, indicating that nitrogen was not inserted into deeper MoS$_2$ layers in this process. Furthermore, using the XPS data acquired at 80° the resulting $\Theta_N$ is ~1 ML, which is consistent with the coverage obtained at 45°.

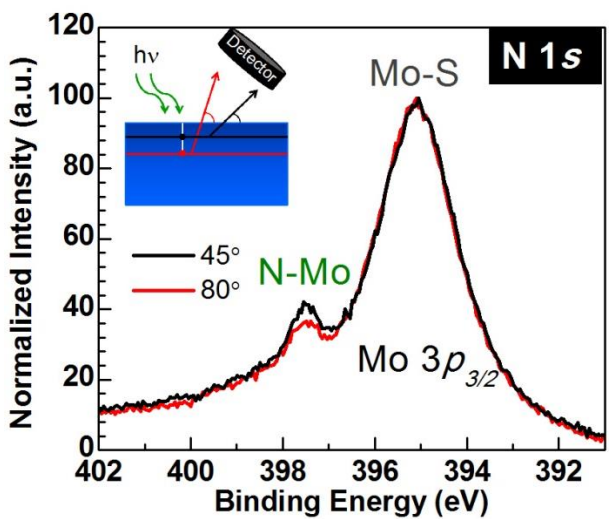

**Figure 6S.** N 1s spectra from 60 min N$_2$ plasma treated MoS$_2$ acquired at take-off angles of 45° (surface sensitive) and 80° (bulk sensitive).

| Take-off Angle | $I_{N\,1s}/I_{Mo\,3d_{5/2}}$ Ratio | $\lambda_{Mo}$ | N coverage (ML) |
|---|---|---|---|
| 45° | 0.081 | 2.431 | 0.99 |
| 80° | 0.066 | 2.493 | 1.06 |

**Table 1S.** Calculated nitrogen coverage obtained at take-off angles of 45° and 80°, and corresponding integrated intensity ratios for N1s and Mo $3d_{52/2}$, and inelastic mean free path λ values used for the calculation.